\documentclass[showpacs,pra,preprint,unsortedaddress,superscriptaddress]{revtex4}
\usepackage{amsmath,amssymb}% Advanced math
\usepackage{graphicx}% Include figure files
\usepackage{dcolumn}% Align table columns on decimal point
\usepackage{bm}% bold math
\begin{document}
\newcommand{\bp}{{\bf p}}
\newcommand{\qq}{{\bf ??}}
\newcommand{\ket}[1]{\left|#1\right\rangle}
\newcommand{\bra}[1]{\left\langle#1\right|}
\newenvironment{alert}{\begingroup\color{red}}{\endgroup}
\newenvironment{marca}{\begingroup\color{blue}}{\endgroup}
\sloppy

\title{Hardy's Test versus the CHSH Test of Quantum Non-Locality: Fundamental
  and Practical Aspects}
\author{Daniel Braun}
\email{braun@irsamc.ups-tlse.fr}
\affiliation{Laboratoire de Physique Th\'eorique,   Universit\'e Toulouse,
  UPS and CNRS, FRANCE}
\author{Mahn-Soo Choi}
\email{choims@korea.ac.kr}
\affiliation{Department of Physics, Korea University, Seoul 136-713, Korea}
\date{\today}
\begin{abstract}
We compare two different tests of quantum non-locality, both in theoretical
terms and with respect to a possible implementation in a mesoscopic circuit:
Hardy's test 
[Hardy, Phys. Rev. Lett. \textbf{68}, 2981 (1992)] 
and the CHSH test, the latter including a recently discovered inequality
relevant for experiments with three possible outcomes 
[Collins and Gisin, J.~Phys.~A \textbf{37}, 1775 (2004)]. 
We clarify the geometry of the correlations defined by Hardy's equations with
respect to the polytope of causal correlations, and show that these equations
generalize to the CHSH inequality if the slightest imperfections in the setup
need to be taken into account.  We propose a mesoscopic circuit consisting of
two interacting Mach-Zehnder interferometers in a Hall bar system for which
both Hardy's test and the CHSH test can be realized with a simple change of
gate voltages, and evaluate the robustness of the two tests in the case of
fluctuating experimental parameters.  The proposed setup is remarkably robust
and should work for fluctuations of beam splitter angles or phases up to the
order of one radian, or single particle loss rates up to about 15\%.
\end{abstract}
% \pacs{03.67.-a, 03.67.Lx, 03.67.Mn}
% 03.65.Ud Entanglement and quantum nonlocality (e.g. EPR paradox, Bell's
% inequalities, GHZ states, etc.) (for entanglement production and
% manipulation, see 03.67.Bg; for entanglement measures, witnesses etc., see
% 03.67.Mn; for entanglement in Bose-Einstein condensates, see 03.75.Gg)
%
% 73.23.-b	Electronic transport in mesoscopic systems
% 73.43.-f	Quantum Hall effects
\maketitle
%\tableofcontents
%\newpage

\section{Introduction}\label{intro} 
The belief that nature should be describable by a local realistic theory is
deeply rooted in our classical intuition \cite{Einstein35}. Impressive
evidence has been accumulated to the contrary, however, starting with the
pioneering work by Bell \cite{Bell64}. He showed that the assumptions of
locality and reality lead, in the framework of classical probability theory,
to bounds on the correlations of measurement outcomes of spatially separated
observers. Experiments with entangled photons have shown that the quantum
world does not obey these bounds \cite{Aspect82}. Violations of Bell's
inequality by tens of standard deviations have been observed meanwhile, while
being in excellent agreement with the quantum mechanical predictions
\cite{Aspect99}. Such quantum correlations have become known as ``quantum
non-locality''. Several proposals have been put forward to observe
entanglement\cite{Beenakker03b,Beenakker04,Samuelsson04a,Lebedev04,Lebedev05,DiLorenzo05}
or even a violation of Bell's inequality in mesoscopic circuits
\cite{Samuelsson03a,Kang07}.
None of these has been implemented experimentally so far,
even though the two-particle Aharonov-Bohm effect\cite{Samuelsson04a}
demonstrated in a
recent experiment\cite{Neder07a}
suggests the presence of the electron entanglement.
On the other hand, a
violation of a modified Bell's inequality has been observed in a circuit--QED
system very recently \cite{Ansmann08}.
From a foundational perspective, it is
desirable to observe a violation of a Bell inequality in a material system, as
one of the key requirements for a conclusive refutation of any local hidden
variable (LHV) description of an experiment is the knowledge of the pair
production rate. As was shown by Santos, an LHV description is always possible
if the probabilities are obtained as relative frequencies normalized to (the
sum of) joint--coincidence rates rather than to absolute pair production
rates, even in the case of ideal polarizers and detectors
\cite{Santos92}. This requirement is hard to meet in quantum optical
experiments, but should be feasible with material particles, which are easier
to keep track of than photons. Beautiful progress was recently achieved in
this direction in an experiment with trapped ions \cite{MatsukevichMMOM08}.

In this paper we compare two different kinds of tests of quantum non--locality
and their mutual relations: Hardy's
test\cite{Hardy92,Hardy93a} and the CHSH test \cite{ClauserHSH69} (the
Bell-type test based on the CHSH inequality).
We compare these tests both on a purely
theoretical level, and in relation to a possible experimental realization
using a specific setup in a mesoscopic circuit, paying particular attention to
the range of parameters in which these tests are expected to signal violation
of non--contextual realism, a generalization of local realism (see below).
The motivation for the first part lies in the fact that Hardy's test, which is
one of the tests of ``non--locality without inequalities'', has always stood
apart from other tests by its simplicity and apparent reliance on pure logic.
Mermin has called it ``the best version of Bell's theorem'' \cite{Mermin95}.
On the other hand, the CHSH inequality is special among all
  Bell-type 
inequalities: It is known to be the only relevant Bell-type inequality (for
bipartite experiments with two observables per observer with two possible
outcomes each) in the sense that a CHSH inequality is always violated if any
other Bell--type inequality is violated (but not necessarily the other way
around) \cite{Fine82a,Fine82b,WernerW01,CollinsG04,BarrettLMPPR05}.
The question then arises, what role Hardy's test plays in this context.  We
discuss the relation between Hardy's test and the CHSH test both in the
absence and  presence of imperfections.  In the former
case, we study the relation between the two tests in geometrical terms,
making obvious 
the connection between the convex sets of joint probabilities of different
measurements involved, while the latter is examined by means of set
theoretical 
arguments \cite{Ghirardi06a,Ghirardi06b,Ghirardi08a}.
We then propose a mesoscopic circuit which allows to implement both Hardy's
test and the CHSH test by a simple change of parameters. This makes it
possible to compare the two tests on an equal footing concerning the range of
parameter fluctuations which, according to QM, would still allow a refutation
of non--contextual hidden variable theories.

\section{Hardy's Test versus The CHSH Test}

\subsection{Local Hidden Variable Theories}
\label{sec:NCHVT}

As pioneered by Bell\cite{Bell64,Bell66a}, the non--locality
of the quantum world can be tested by 
attempting to construct a
local 
hidden variable theory that reproduces all quantum mechanical
predictions.  Such an approach leads to necessary conditions for LHV
theories in the form of bounds on the correlations, the famous Bell
inequalities. If in an experiment a violation of such an inequality is
observed, a LHV description is ruled out 
and the non--locality of the quantum world is established.

Let us consider a possible local hidden-variable description of a
bipartite system.  LHV models are defined by the joint probabilities
$P(m_1,m_2|M_1,M_2)$ to get a pair of outcomes $m_1$ and $m_2$ for the
observables $M_1$ and $M_2$ on ``particle'' 1 and 2, respectively.  There can
be $n_1$ ($n_2$) observables and $k_1$ ($k_2$) outcomes per observable on the
side of ``particle'' $1$ ($2$).  We suppose that the particle $1$ ($2$) is
possessed by Alice (Bob).

There are overall $d=n_1n_2k_1k_2$ possible joint probabilities, but they are
not all independent and have to satisfy several constraints.  First of all,
all joint probabilities must be non--negative and normalized, such that for
any measurement setting $(M_1,M_2)$
\begin{equation} \label{Cn}
\sum_{m_1,m_2}P(m_1,m_2|M_1,M_2)=1\,.
\end{equation}
This implies that 
\begin{equation} \label{Cpos}
0\le P(m_1,m_2|M_1,M_2)\le 1
\end{equation}
for all $m_1,m_2,M_1,M_2$.
Secondly, the joint probabilities must respect the causality constraint (also
called non--signaling constraint). This means, that the reduced probabilities
on either side must not depend on the measurement settings on the other side
--- otherwise superluminal signaling would be possible over sufficiently large
distances:
\begin{subequations}
\label{Cns}
\begin{align}
\sum_{m_2}P(m_1,m_2|M_1,M_2)&= \sum_{m_2'}P(m_1,m_2'|M_1,M_2')
\quad\forall m_1,M_1,M_2,M_2' \,,\\
\sum_{m_1}P(m_1,m_2|M_1,M_2)&= \sum_{m_1'}P(m_1',m_2|M_1',M_2)
\quad\forall m_2,M_2,M_1,M_1' \,.
\end{align}
\end{subequations}
Bell-type correlations are further constrained by the request of
locality for each value of the hidden variables $\lambda$. In a LHV
theory, the outcome of {\em any single run} must be, for both observers, a
function of the hidden variables $\lambda$ and the local measurement setting
alone, such that the joint probabilities are given by
\begin{equation} \label{Cnchv}
P(m_1,m_2|M_1,M_2)
= \sum_\lambda p_\lambda P^{(1)}_\lambda(m_1|M_1)P^{(2)}_\lambda(m_2|M_2)\,,
\end{equation}
where $p_\lambda$ is the normalized distribution of hidden variables, and
$P^{(j)}_\lambda(m_j|M_j)$ is the probability for a given hidden variable
$\lambda$ that the measurement result is $m_j$ if an observable $M_j$ of particle $j$ ($j=1,2$) is 
measured.
We have written down the
locality constraint
for a stochastic hidden variable theory. A deterministic HV theory is a
special case, where all probabilities $P^{(1)}_\lambda(m_1|M_1)$ and
$P^{(2)}_\lambda(m_2|M_2)$ are either zero or one. On the other hand, any
stochastic HV theory can be made deterministic
\cite{Fine82a,Fine82b,WernerW01}.
Note that the causality constraint above works on the level of the actually
observed probabilities (i.e.~for averages over hidden variables), whereas the
locality constraint
(\ref{Cnchv}) is based on the request of non--signaling for each 
value of the 
hidden variables.

\subsection{Ideal Hardy's Test}

In 1992, Hardy proposed a novel experiment, which allows to test whether
Nature can be described by a local 
realistic theory \cite{Hardy92}.  The experiment can be based on any pair of
observables $M_1=X_1,Y_1$ and $M_2=X_2,Y_2$ with two mutually exclusive
outcomes (which we will take as $\pm 1$ for concreteness) for each observable.
Alice (Bob) has the free choice to measure either $X_1$ or $Y_1$ ($X_2$ or
$Y_2$) in any run of the experiment. The measurements of Alice and Bob should
be space-like separated, such that the measurement settings of Alice will, if
one accepts Einstein locality, not influence the outcomes of Bob's experiment
and vice versa. 
The situation first analyzed by Hardy \cite{Hardy92} assumes three
vanishing joint-probabilities,
\begin{align}
P({+,+}|X_1,X_2)&=0\label{l1}\,,\\
P({+,-}|Y_1,X_2)&=0\label{l2}\,,\\
P({-,+}|X_1,Y_2)&=0\label{l3}\,,
\end{align}
Suppose an experimental setup can be found where these three equations are
fulfilled. Then, if Nature can be described by a LHV theory, it
follows immediately that also  
\begin{equation} \label{l4}
P(+,+|Y_1,Y_2)=0
\end{equation}
must be satisfied.  In the following we will call Eqs.~(\ref{l1})-(\ref{l4})
``Hardy's equations''.
To see how a violation of (\ref{l4}) under given assumptions
(\ref{l1})--(\ref{l3}) implies non--locality, suppose that an event with
$y_1=y_2=+1$ was detected for the simultaneous measurement of $Y_1$ and $Y_2$.
It follows from Eq.~(\ref{l2}) that on Bob's side the outcome $x_2$ of the
measurement of $X_2$ must have
had the value $+1$, as $y_1=1$ can never appear with $x_2=-1$, and $x_2=+1$ is
the only alternative. Similarly, from Eq.~(\ref{l3}) follows that $x_1$ must
have had the value $+1$, as $y_2=1$ can never appear with $x_1=-1$, and
$x_1=+1$ is the only alternative. Furthermore, due to the locality
assumption, the value of $x_2$ cannot depend on whether Alice measured $x_1$
or $y_1$, and $x_1$ cannot depend on whether Bob measured $x_2$ or $y_2$. So
one concludes in a LHV theory that {\em both} $x_1$ and $x_2$ must have had
the values $+1$. This, however, is excluded by Eq.~(\ref{l1}). As a
consequence, if Eqs.~(\ref{l1})--(\ref{l3}) are fulfilled, even a single event
$y_1=y_2=+1$ amidst any finite series of measurements rules out that
the experiment can be described by a local realistic  
theory.

Hardy managed to construct a pure state for which, according to QM, just this
happens \cite{Hardy92}.  Later, his argument was generalized and it was shown
that almost any pure state of any system with an arbitrary number of particles
and arbitrary dimension of Hilbert space can be used
\cite{Hardy93,Goldstein94,Ghirardi05}, and even a large class of mixed states
\cite{Ghirardi06a}.

It is instructive to determine how strongly Eq.~(\ref{l4}) can be violated
according to QM. Following Ref.\cite{Goldstein94}, suppose a pure state $|
\psi \rangle=b|01\rangle+c|10\rangle+d|11\rangle$ is prepared (normalized to
$|b|^2+|c|^2+|d|^2=1$), where e.g.~$|01\rangle=|0\rangle\otimes|1\rangle$, the
first state belongs to Alice, the second to Bob, and $|0\rangle$ and
$|1\rangle$ are orthogonal basis states, which we will take as computational
basis. In fact, any pure two--qubit state which is neither a product state nor
a maximally entangled state can be brought to this form through an appropriate
choice of orthogonal basis states \cite{Goldstein94}. Assume furthermore that
the measurement operators on Alice's side are defined by $X_1=|0\rangle\langle
0|-|1\rangle\langle 1|$ and $Y_1=|y_1^+\rangle\langle
y_1^+|-|y_1^-\rangle\langle y_1^-|$, where
\begin{equation}
|y_1^+\rangle=\frac{d^*|0\rangle-b^*|1\rangle}{\sqrt{|b|^2+|d|^2}}
\end{equation}
and
\begin{equation}
|y_1^-\rangle= \frac{b|0\rangle+d|1\rangle}{\sqrt{|b|^2+|d|^2}}
\end{equation}
The possible
measurement outcomes are obviously $\pm 1$ for both measurements. Similarly,
for Bob we define $X_2=|0\rangle\langle 0|-|1\rangle\langle 1|$ and
$Y_2=|y_1^+\rangle\langle y_2^+|-|y_2^-\rangle\langle y_2^-|$,
where
\begin{equation}
|y_2^+\rangle=\frac{d^*|0\rangle-c^*|1\rangle}{\sqrt{|c|^2+|d|^2}}
\end{equation}
and
\begin{equation}
|y_2^-\rangle=\frac{c|0\rangle+d|1\rangle}{\sqrt{|c|^2+|d|^2}}\,.
\end{equation}
It is then straightforward to verify \cite{Goldstein94} that
Eqs.~(\ref{l1})--(\ref{l3}) are fulfilled, whereas
\begin{equation} \label{Py1y2id}
P({++}|YY)=\frac{|bcd|^2}{(|b|^2+|d|^2)(|c|^2+|d|^2)}
\neq 0 \,.
\end{equation}
where we have omitted the particle indices $1$ and $2$ (hereafter we will
adopt this convention unless there is a risk of confusion).
The joint probability $P({++}|YY)$ is maximized for $|b|=|c|$ and
$|d|=\sqrt{5}-2$, in which case
\begin{equation} \label{PHid}
P({++}|YY)=\frac{5\sqrt{5}-11}{2} \simeq 0.09017\,.
\end{equation}
Thus,
 about 9\% of the experimental outcomes should falsify any LHV theory, and
 this is the largest possible value \cite{Goldstein94}.  We
 call such an optimized state 
\begin{equation} \label{psiIH}
|\psi\rangle=\sqrt{2\left(\sqrt{5}-2\right)}
\left(|01\rangle+|10\rangle\right)+(\sqrt{5}-2)|11\rangle
\end{equation}
an ``ideal Hardy state''.

``Hardy non--locality'' has so far stood apart from other tests of local
realistic theories in several aspects. Firstly, it does not rely on
inequalities, but apparently on pure logic. In the ideal setting proposed by
Hardy, a single measurement 
event can invalidate all LHV theories. Note that for three particles such
tests are known and can be based, e.g.~on the GHZ state
\cite{Greenberger89}. Secondly, Hardy non--locality does \emph{not} work for
maximally entangled states, in contrast to the standard Bell's inequality. It
is well known that the CHSH inequality \cite{ClauserHSH69} is maximally
violated for a singlet state (and appropriately chosen measurements).
It may be for these reasons that Hardy non--locality has been called ``the
best version of Bell's theorem'' \cite{Mermin95}.  Substantial efforts have
been spent to observe Hardy non--locality experimentally. An experiment with
photons was performed by Bouwmeester et al.\cite{Irvine05} using the bunching
of photons at a beam splitter (BS) in order to create a Hardy state, and by Di
Giuseppe et al.~using polarized photons in a non--maximally entangled state
\cite{Giuseppe97a}.

\subsection{The Geometry of Hardy's Test}

Above we have seen several apparent differences between Hardy's test and the
CHSH test.
But what is the precise
relationship between Hardy's test and the CHSH test?  To answer this
question, we found it very instructive to investigate the geometry of the
sets of joint probabilities defined by the two tests.

Given a set of observables $M_1=X_1,Y_1$ and $M_2=X_2,Y_2$ with outcomes
$x_j=\pm$ and $y_j=\pm$ ($j=1,2$), an entire correlation table of 16 joint
probabilities $P(m_1,m_2|M_1,M_2)$ can be regarded as a point in the
16-dimensional vector space $\mathbb{R}^{16}$.
The correlation tables cannot span the whole space $\mathbb{R}^{16}$
because of the various constraints imposed on the joint probabilities, as
discussed in Section~\ref{sec:NCHVT}.  For example, the positivity and
normalization constraints, (\ref{Cn}) and (\ref{Cpos}), define a convex
subset.  The causality constraints~(\ref{Cns}) restrict further the subset,
and lead to a convex subset in the form of polytope which we call the
``causal polytope'' $\mathcal{C}$.  A polytope is the higher dimensional
generalization of a polyhedron in three dimensions with flat surfaces
(i.e.~given by linear equations) and a finite number of vertices. For the
situation considered here, there are 12 linear 
equations, 4 from the normalization constraints~(\ref{Cn}) and 8 from the
causality constraints~(\ref{Cns}).  Only 8 of them are linearly independent,
and thus the causal polytope $\mathcal{C}$ is 8 dimensional (8D).  It has 7D 
facets.
Correlation tables restricted additionally to (\ref{Cnchv}) form also a
convex polytope, $\mathcal{L}$, which is commonly called the
``local polytope``, or ``Bell polytope''. 
The local polytope $\mathcal{L}$
lies inside the causal polytope $\mathcal{C}$
\cite{Fine82a,Fine82b,WernerW01,BarrettLMPPR05}. As was shown by Fine, the
CHSH inequalities together with the positivity constraints~(\ref{Cpos}) form
all the facets of $\mathcal{L}$, and give therefore a complete
characterization of all LHV correlations. Quantum correlations can lie
outside $\mathcal{L}$, but are always inside $\mathcal{C}$. They also form a
convex set, but not in the form of a polytope. They are still restricted by
Cirel'son's bound, which gives an upper bound $2\sqrt{2}$ for the left hand
side of the CHSH inequality \cite{Cirelson80}.

All vertices of the polytope $\mathcal{C}$ have all joint probabilities either
equal zero or one. There are 24 vertices, 16 of which represent local
correlations (called ``local vertices''), and 8 represent non--local
correlations. The vertices which represent local correlations are the
vertices of the Bell polytope. They can be parameterized by
four binary 
variables, $\alpha,\beta,\gamma,\delta\in \{0,1\}$. If we also code
measurement outcomes and observables with binaries ($M_j=X_j,Y_j\mapsto 0,1$
and $m_j=\pm\mapsto 0,1$ for $j=1,2$), they can be found from
\cite{BarrettLMPPR05}
\begin{equation} \label{lver}
P(m_1,m_2|M_1,M_2) =
\begin{cases}
1 \,,& m_1=\alpha M_1\oplus \beta \mbox{ and }
m_2=\gamma M_2\oplus
\delta\\
0 \,,& \mbox{otherwise}\,,
\end{cases} 
\end{equation}
where $\oplus$ denotes addition modulo 2. Hardy's equations
(\ref{l1})--(\ref{l4}) are four more independent linear constraints,
which thus restrict us to a 4D subspace of the 8D polytope $\mathcal{C}$.
Indeed, the remaining joint probabilities [besides the ones
chosen zero in Eqs.~(\ref{l1})--(\ref{l4})], can be parameterized as
\begin{equation}
\label{Hardy4D}
\begin{split}
P({+-}|XX)&=1-P({--}|XY) \,,\\
P({-+}|XX)&=1-P({--}|YX) \,,\\
P({--}|XX)&=-1+P({--}|XY)+P({--}|YX) \,,\\
P({+-}|XY)&=1-P({--}|XY)-P({-+}|YY)\,,\\
P({++}|XY)&=P({-+}|YY)\,,\\
P({++}|YX)&=1-P({--}|YY)+P({-+}|YY) \,,\\
P({-+}|YX)&=-P({--}|YX)+P({--}|YY)+P({-+}|YY)\,,\\
P({+-}|YY)&=1-P({--}|YY)-P({-+}|YY) \,.
\end{split}
\end{equation}
All 16 joint probabilities are now determined once we specify $P({--}|XY)$,
$P({--}|YX)$, $P({-+}|YY)$, and $P({--}|YY)$.
Thus, Hardy's equations span a 4D polytope $\mathcal{H}$ which we will call
``Hardy's polytope''.
In the following we will group the four independent joint probabilities in a
4D vector ${\bf p}=(P({--}|XY), P({--}|YX), P({-+}|YY), P({--}|YY))$. It is
straightforward to check that five of the 16 local vertices
satisfy Eq.~(\ref{Hardy4D}), namely those given by $\bp=(1,0,0,0)$,
$\bp=(1,0,0,1)$, $\bp=(1,1,0,1)$, $\bp=(0,1,0,1)$, and $\bp=(0,1,1,0)$. The
other local vertices can be covered by another set of Hardy's
equations generated from (\ref{l1})--(\ref{l4}) through local
permutations of measurements, but in the following we will focus on one given
set of Hardy's equations, i.e.~on a single polytope $\mathcal{H}$.
The five local vertices are the \emph{only} vertices of
$\mathcal{H}$, as can be directly verified by looking at all combinations of
probabilities equal zero or one for the elements of $\bp$, calculating the
remaining probabilities, and checking whether they fulfill normalization,
positivity, and causality. Taking one of the five vertices as origin, we have
four linearly independent vectors pointing to the other four vertices, which
thus allow to entirely span $\mathcal{H}$. This by itself does not mean yet
that $\mathcal{H}$ lies in a facet of $\mathcal{L}$, as a polytope spanned by
local vertices might lie in the interior of
$\mathcal{L}$. However, we know that for all points within $\mathcal{H}$,
i.e.~in particular with the three probabilities in (\ref{l1})--(\ref{l3})
equal zero, adding an infinitesimally small value to $P({++}|YY)$ moves us
outside the realm of LHV theories, by construction of Hardy's argument. Thus,
for all points in $\mathcal{H}$ where a positive $P({++}|YY)$ is not prohibited
by other constraints (normalization and positivity of other joint
probabilities) $\mathcal{H}$ must lie inside an \emph{interface} between
$\mathcal{L}$ and the remainder of $\mathcal{C}$. This interface must be a
single facet of $\mathcal{L}$ or an edge, as otherwise $\mathcal{H}$ would not be 
convex. (For points where $P({++}|YY)$ is prevented from taking positive values,
$\mathcal{H}$ may lie within one of the trivial facets of $\mathcal{L}$ given
by the positivity and upper bound one of all joint--probabilities).
Thus, Hardy's equations define a 4D polytope contained within a 7D facet,
which forms a boundary of the local polytope $\mathcal{L}$ with
the set of non--local correlations, and
hence corresponds to a CHSH inequality.

Suppose now that we add a small value $\epsilon$ to any of the probabilities
in the first three Hardy equations, Eqs.~(\ref{l1})--(\ref{l3}), i.e.~move
outside of $\mathcal{H}$ in a different direction. Since we were already
inside the interface between the local and the other causal
correlations described by CHSH inequalities, the best we can do if we want to
stay in $\mathcal{L}$ is to move within the facet. This suggests that a
generalization of Hardy's equations to a necessary condition for LHV theories
with finite values of the probabilities in Eqs.~(\ref{l1})--(\ref{l3}) should
lead immediately to a CHSH inequality. Indeed, we will see in the following
subsection (Section~\ref{sec.imp}) that this is the case.

\begin{figure}
\centering
\includegraphics[width=4cm]{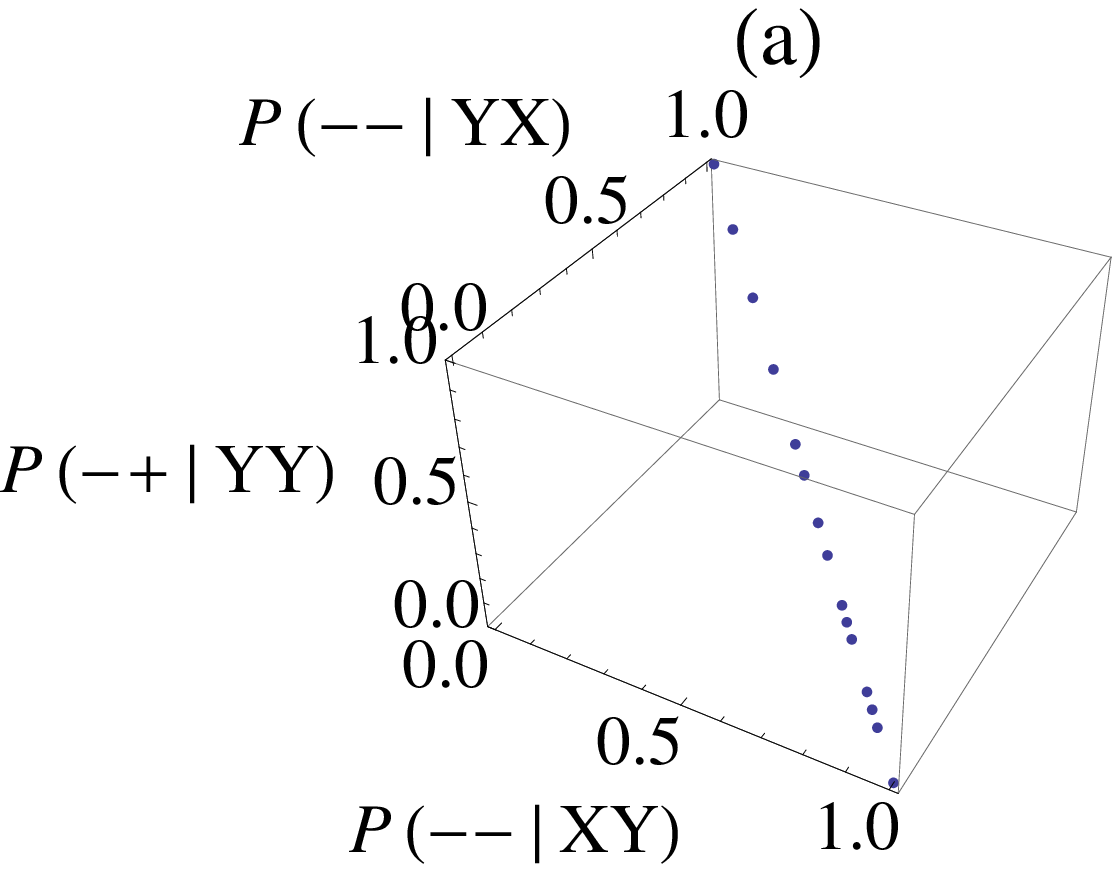}
\includegraphics[width=4cm]{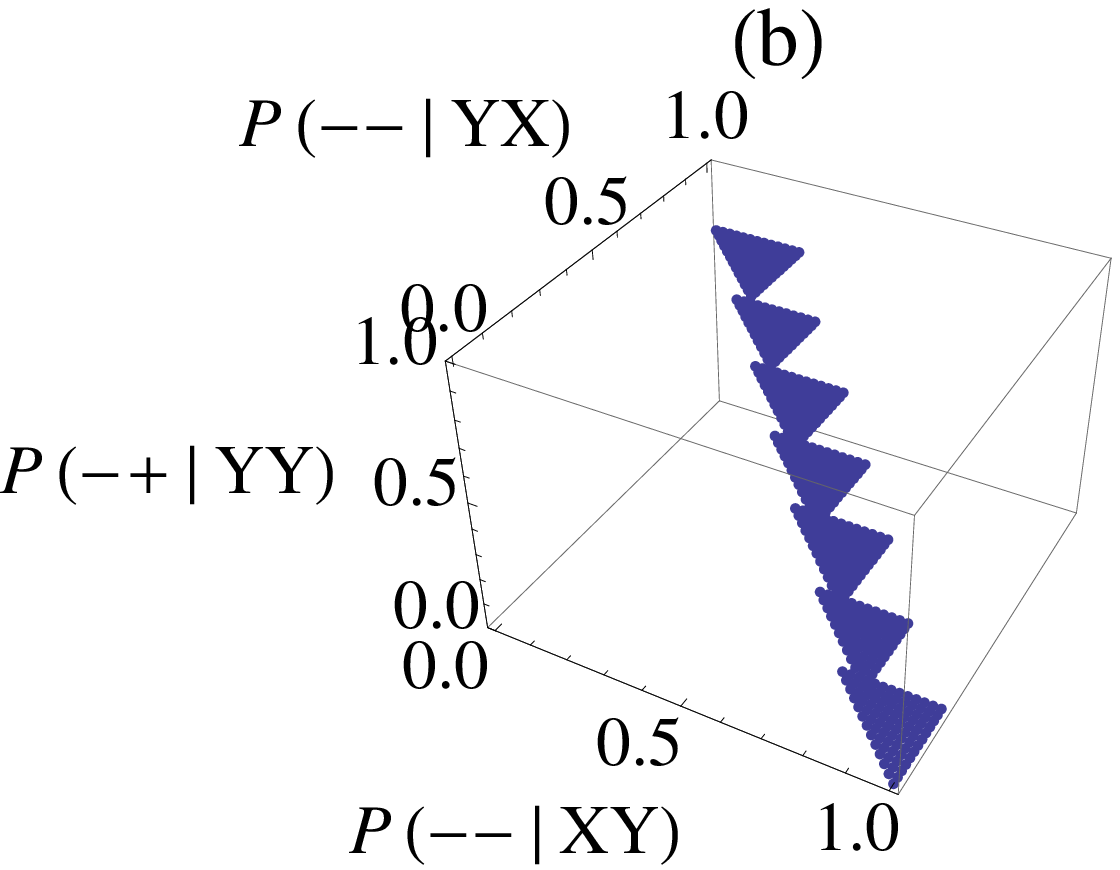}
\includegraphics[width=4cm]{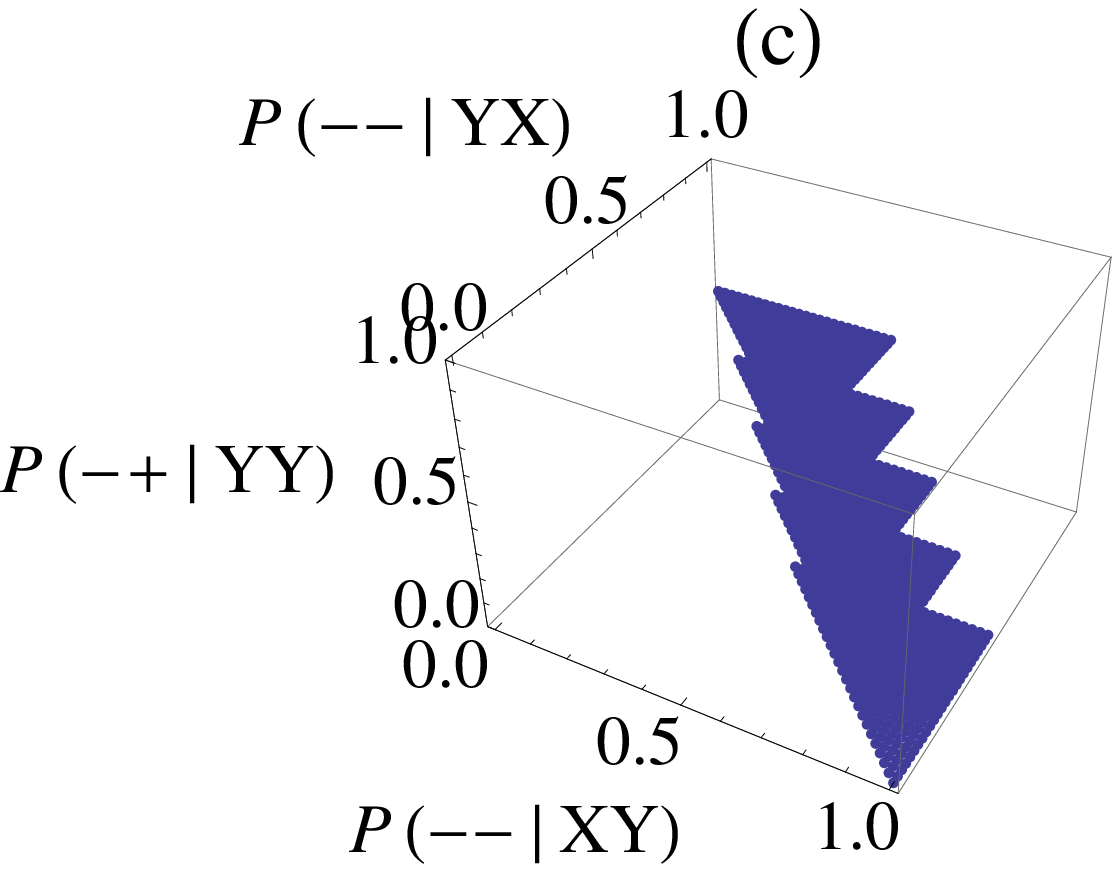}
\includegraphics[width=4cm]{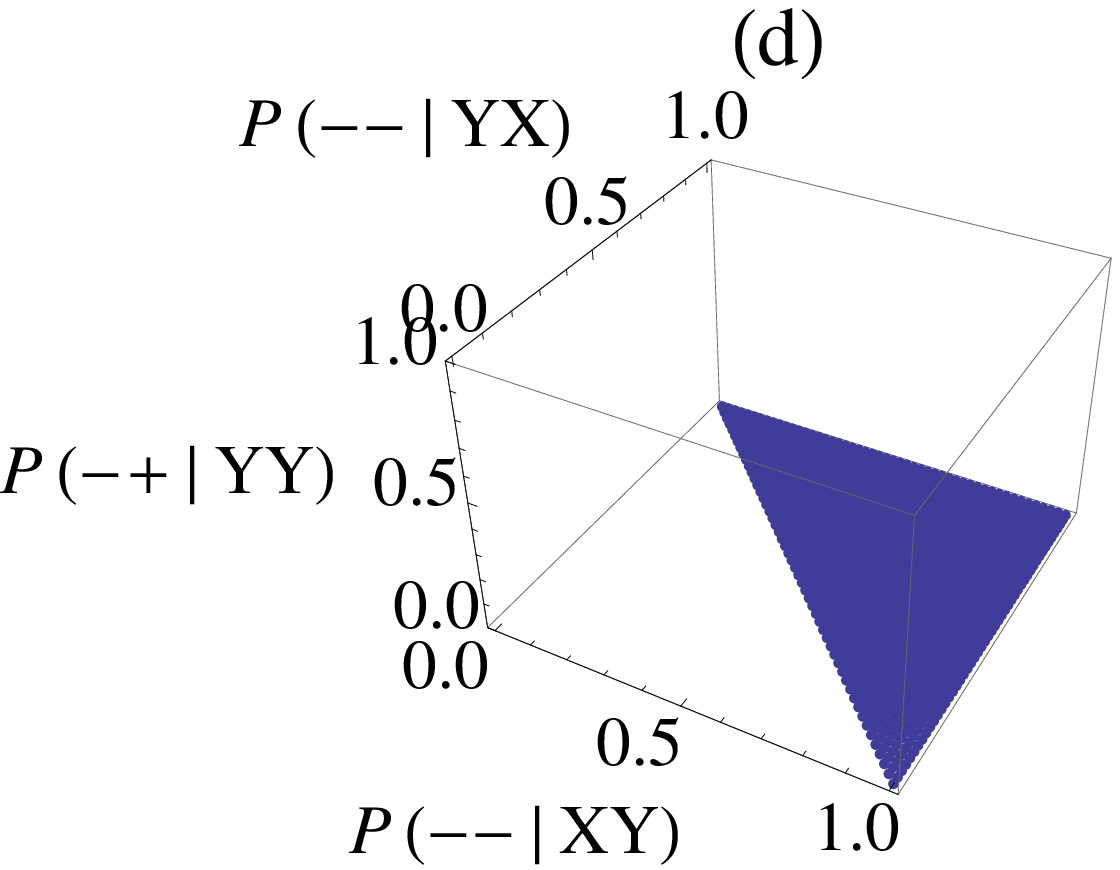}
\caption{3D cross--sections of the Hardy--polytope $\mathcal{H}$ for
  $P({--}|YY)=0$ (a), 0.25 (b), 0.5 (c), and 1 (d). All cross--sections are
  cut once more at constant values $P({-+}|YY)$ to reveal the triangular 2D
  cross-sections.}
\label{fig.Hp}
\end{figure}

In order to visualize the polytope $\mathcal{H}$ in the subspace of the
components of $\bp$, we present in Fig.\ref{fig.Hp} several 3D cuts of
$\mathcal{H}$, namely for $P({--}|YY)=0,\frac{1}{4},\frac{1}{2}$, and 1. We see
that for $P({--}|YY)=0$ the polytope degenerates to a straight diagonal line,
from $\bp=(0,1,1,0)$ to $\bp=(1,1,0,0)$. For finite values of $P({--}|YY)$,
the line widens to a cylinder with the cross-section of an right--angled
isosceles triangle in the planes of constant $P({-+}|YY)$. These triangles move
along the mentioned diagonal with increasing $P({-+}|YY)$, till they hit the
boundary $P({--}|XY)=0$. The length of the short sides of the triangles are
given by $P({--}|YY)$. Altogether, we can describe the polytope $\mathcal{H}$
by the four inequalities
\begin{eqnarray}
0&\le& P({--}|YY)\le 1\,,\\
0&\le& P({-+}|YY)\le 1-P({--}|YY)\,,\\
1-P({-+}|YY)-P({--}|YY)&\le& P({--}|XY)\le 1-P({-+}|YY)\,,\\
1-P({--}|XY)&\le& P({--}|YX)\le 1-P({--}|XY)+P({--}|YY)\,.
\end{eqnarray}
The five local vertices which satisfy (\ref{Hardy4D}) show up here
as corners $(1,0,0)$ (twice, once for $P({--}|YY)=0$ and once for
$P({--}|YY)=1$), $\bp=(1,1,0)$ and $\bp=(0,1,0)$ for $P({--}|YY)=1$, and
$\bp=(0,1,1)$ for $P({--}|YY)=0$.

A final remark is in order about the quantum states which allow to falsify
LHV theories using Hardy's test. As
mentioned, almost all pure states allow to demonstrate Hardy non--locality,
but the singlet state, which 
violates the CHSH inequality maximally, does not. How is this possible if
$\mathcal{H}$ is a subset of an interface described by a CHSH inequality?
The reason for this is 
that no set of observables $X_1,Y_1,X_2,Y_2$
can be found such that Eq.~(\ref{l4}) is violated while Eqs.~(\ref{l1}),
(\ref{l2}), and 
(\ref{l3}) are all satisfied.  The set of
correlations which satisfy Hardy's equations  (\ref{l1}),
(\ref{l2}), and 
(\ref{l3}) define a 5D subset of $\mathcal{C}$, and only in this subset can
Hardy's test exclude LHV theories. The singlet state leads to correlations,
which, no matter the choice of observables, are outside of this subset.

\subsection{Hardy's Test in the Presence of Imperfections}
\label{sec:imperfection}

In any realistic experimental situation, it will be difficult to fulfill
Eqs.~(\ref{l1})--(\ref{l3}) \emph{exactly}.  It is therefore essential to
analyze the effects of various imperfections, which may lead to finite values
of the probabilities in Eqs.~(\ref{l1})--(\ref{l3}).  Errors can occur in the
preparation of the ideal quantum mechanical state $\ket\psi$, in the
construction of the measurement operators $X_j$ and $Y_j$, and due to
detection problems, including particle loss. Suppose then that, say,
$P({++}|XX)$ has some finite but small value,
$P({++}|XX)=\epsilon_1$. Immediately the logic of Hardy's argument
ceases to work, and nothing prevents an outcome ($++|YY$). The same
holds true for the other two probabilities in Eqs.~(\ref{l2}-\ref{l3}), for
which we may assume that they take on finite values
$P({+-}|YX)=\epsilon_2$, and $P({-+}|XY)=\epsilon_3$. Furthermore,
the logic of Hardy's argument also ceases to work if a particle can be lost,
as this corresponds to a \emph{third} outcome, which we will label `0' for any
measurement. We should therefore also consider the situation where
probabilities such as $P(+0|YX)=\epsilon_4$ and
$P(0+|XY)=\epsilon_5$ can become finite.  For continuity reasons it is
clear that $P({++}|YY)$ cannot jump immediately to arbitrarily large
values if any of the $\epsilon_i$ takes on a very small but finite value. In
other words, there should be a bound on $P({++}|YY)$ depending on the
$\epsilon_i$. We will now show that this bound is equivalent to the CHSH
inequality.

The key to generalizing Hardy's argument to finite values of the
$\epsilon_i$ is replacing logical implications by set--theoretical
inclusions. This approach was pioneered very recently by Ghirardi and
Marinatto (GM) 
\cite{Ghirardi06a,Ghirardi06b}. The sets in
question are sets of values of hidden variables which imply certain outcomes
of measurements. 
Following the
 steps in  
 \cite{Ghirardi06a}, only slightly generalized to different $\epsilon_\nu$
 and equalities instead of bounds for the joint probabilities, we
 immediately find the necessary condition  
\begin{equation}
\label{ineqd}
P({++}|YY)\le\sum_{\nu=1}^5\epsilon_\nu
\end{equation}
for any LHV theory. 
Note that (\ref{ineqd}) is based solely on
classical set theory and does not make any assumption on how the measurement
outcomes are generated.
The probabilities $\epsilon_4$ and $\epsilon_5$  may not
 be measurable through a  
direct correlation measurement, but conservation of probability demands
\begin{equation}
P({+0}|YX)=P^{(1)}({+}|Y)-P({++}|YX)-P({+-}|YX)
\end{equation}
and similarly for $P({0+}|XY)$. Using this and inequality (\ref{ineqd}),
we are immediately led to the CH inequality \cite{ClauserH74},
\begin{multline} \label{CHSHp}
P({++}|XY)+P({++}|YX)+P({++}|YY) \\{}
-P({++}|XX)-P^{(1)}({+}|Y)-P^{(2)}({+}|Y) \le 0\,.
\end{multline}
The CH inequality (\ref{CHSHp}) is mathematically equivalent
to the 
more familiar CHSH inequality \cite{ClauserHSH69} based on expectation
values for dichotomic 
observables with measurement outcomes $\pm 1$, 
\begin{equation} \label{CHSHe}
\langle X_1Y_2 \rangle+\langle Y_1X_2 \rangle+\langle X_1X_2 \rangle-\langle
Y_1Y_2 \rangle-2\le 0 \,,
\end{equation}
if particle loss is excluded \cite{Nielsen00}. This can
be easily seen by using conservation of probability to express all four
probabilities appearing in an expectation value like $\langle Y_1 X_2\rangle$
in terms of the sole probability $P({++}|YX)$ (and similarly for all
other expectation values).  
Thus, the CHSH inequality can be considered as a natural generalization of
Hardy's equations --- or Hardy's equations as a special case of the CHSH
inequality --- once the slightest imperfections need to be taken into account
(see also \cite{Ghirardi08a}).  Not withstanding the
fact that Hardy's test has always been 
considered apart from other quantum non--locality tests, this should come to
no surprise, as in the 2222 scenario (i.e., $n_1=n_2=k_1=k_2=2$; two
observables for both Alice and Bob with two possible values each), the only
relevant inequality is the CHSH inequality
\cite{Fine82a,Fine82b,CollinsG04,WernerW01}, in the sense that if any
inequality linear in the relevant joint-- and single--particle probabilities
is violated in this scenario, so is one of the CH inequalities constructed
from (\ref{CHSHp}) through symmetry operations such as particle exchange or
relabeling of measurement results. Thus, as soon as Hardy's test needs to be
formulated using bounds on probabilities, the resulting inequality can be at
most as strong as the CHSH inequality.  Note that in the experimental
realization in Ref.\cite{Irvine05} a similar inequality was derived in order
to deal with imperfections.

Our derivation of the CHSH inequality as
generalized Hardy's test also sheds a new light on the former in the following
sense: The inequality (\ref{CHSHp}) provides a
necessary condition for LHV theories 
even if particles can be lost. However, this does not change the status of the
problem of the detector loophole. A violation of (\ref{CHSHp}) would always
have been considered as falsification of a LHV description, if the
probabilities were the ones describing the whole ensemble of pairs used, and
not just the pairs which were detected. Otherwise an additional fair sampling
assumption comes in, which is the origin of the detector loophole.

\subsection{The CHSH Test with Particle Loss}
\label{sec:particle-loss}

In reality, particles can be lost without any measurement
signal, on the way from the source to the detector, or due to
non--ideal detectors.   Particle loss may be considered as a third
measurement outcome, say, `0'.  We are therefore dealing 
with the case of $n_1=n_2=2$ and $k_1=k_2=3$.  In such a
case there exists an inequality, called $I_{2233}$ inequality, 
\begin{eqnarray}\label{I2233}
I_{2233}&=&P({-+}|YY)+P({++}|YY)+P({--}|YY)\nonumber\\ 
&&+P({++}|XY)+P({--}|XY)+P({+-}|XY)\nonumber\\ 
&&+P({++}|YX)+P({--}|YX)+P({+-}|YX)\nonumber\\
&&-P({++}|XX)-P({--}|XX)-P({+-}|XX)\nonumber\\
&&-P^{(1)}({-}|Y)-P^{(1)}({+}|Y)-P^{(2)}({+}|Y)-P^{(2)}({-}|Y)\le 0\,,  
\end{eqnarray}
which is more
relevant than the CHSH inequality
\cite{CollinsGLM02,KaszlikowskiKCZ02}.  In other words, it can detect quantum
non--local correlations, even if the CHSH 
inequality fails to do so. It appears therefore to be
worthwhile examining, whether experiments which include the possibility of
particle loss would not better test for non--locality using the $I_{2233}$
inequality. It is the purpose of the present subsection to show that this is
not the case. 

Interestingly, the inequality is maximally violated
by a non--maximally entangled state, if the three outcomes correspond to
actual quantum states \cite{AcinDGL02}.  However, in the case of particle loss
as third outcome, the events are not independent. If we assume that an
electron is lost with probability $r$ on Alice's side, and with the same
probability (and independently) on Bob's side, $I_{2233}$ becomes
\begin{equation}
\label{I2233r}
\begin{split}
I_{2233}(r)={}&\Big[
P({-+}|YY)+P({++}|YY)+P({--}|YY)\\&{}
+P({++}|XY)+P({--}|XY)+P({+-}|XY)\\&{}
+P({++}|YX)+P({--}|YX)+P({+-}|YX)\\&{}
-P({++}|XX)-P({--}|XX)-P({+-}|XX)\Big](1-r)^2\\&{}
-\Big[P^{(1)}({-}|Y)+P^{(1)}({+}|Y)+P^{(2)}({+}|Y)+P^{(2)}({-}|Y)
\Big](1-r)\le 0 \,.
\end{split}
\end{equation}
Whereas in (\ref{I2233}) the probabilities mean the actually observed ones,
the probabilities in (\ref{I2233r}) are the ideal probabilities without
particle loss.
The latter permit only two values for each observable, and we have
conservation of these ideal probabilities, such as
\begin{equation}
P^{(1)}({+}|X)=P({++}|XX)+P({+-}|XX)
\end{equation}
This allows to rewrite $I_{2233}(r)$
as
\begin{equation}
\label{2233r2}
I_{2233}(r)
=I_{CHSH}(r)-r(1-r)\left[P^{(1)}({+}|X)+P^{(2)}({+}|X)\right]\le 0\,,
\end{equation}
where
\begin{multline}
\label{CHSHr}    
I_{CHSH}(r)=
\Big[P({++}|YY)+P({++}|XY)+P({++}|YX)
-P({++}|XX)\Big](1-r)^2 \\{}
-\Big[P^{(1)}({+}|Y)+P^{(2)}({+}|Y)\Big](1-r) \le 0
\end{multline}
is the CHSH inequality modified for the possibility of particle loss (see also
Section~\ref{sec.imp} below).
For $r=0$ we have $I_{2233}(0)=I_{CHSH}(0)=I_{CHSH}$, as it should be. Due to
the positivity of the term $P^{(1)}({+}|X)+P^{(2)}({+}|X)$ in
(\ref{2233r2}), we 
have $I_{2233}(r)\le I_{CHSH}(r)$, i.e.~if $I_{2233}(r)\le 0$ is violated, so
is $I_{CHSH}(r)\le 0$, but not necessarily the other way round.
Thus, we have established that if the third measurement outcome is particle
loss, the CHSH inequality is more relevant than the $I_{2233}$ inequality,
contrary to the case of genuine three--outcome measurements.

To summarize, we have established the CH inequality (\ref{CHSHp}) (or,
equivalently, the CHSH inequality (\ref{CHSHe}) as the
relevant inequality for all three tests considered above: ideal Hardy's test,
Hardy's test with imperfections, and the CHSH test including particle
loss. This  
leaves open the question, however, which test will be violated in a more
robust way in an experiment. Optimizing different tests leads indeed to
\emph{different} experiments, i.e.~not only the state to be constructed is
different, but so are the corresponding measurement operators.
Since all tests will be based on the CHSH inequality, we will distinguish
different tests by the specific experimental situations.
Under the ``CHSH test'' we understand an experiment using the singlet state
(or, in the case of noise, a state close to the singlet) in order to show a
violation of the CHSH inequality. We will call ``Hardy's test'' an experiment
using a state close to the ideal Hardy state (\ref{psiIH}) in order to show a
violation of the CHSH inequality (see Section~\ref{sec:imperfection}
above). Experimentally, it is desirable to get as strong a violation as
possible, but also in a range of parameters as wide as possible. Since the
CHSH inequality is known to be violated maximally for a singlet state
\cite{Cirelson80}, the CHSH test wins in the first category. However, the
second category is important in the case of uncontrolled fluctuations of
parameters in the experiment, and the question which test fares best in this
category is \emph{a priori} open. It is best answered for a specific
experimental setup, and this is what we are going to discuss now.

\section{Proposed Experimental Setup}
We propose a flexible mesoscopic circuit which allows to perform Hardy's
test and the CHSH test discussed in the previous section, and compare then
the expected 
parameter regions in which these tests are expected to fail according to
quantum mechanics. Space-like separation of measurements is unlikely to
be achieved on a chip in 
the near future, where the measurement stations are separated by a few
$\mu$m. However, the locality condition is known to be a special case of
the more general concept of non--contextuality
\cite{Mermin93}. Non--contextuality in quantum mechanics (QM) means that the
measurement of an observable $A$ does not influence the outcome of the
measurement of another observable $B$ that commutes with $A$. The theories
ruled out by a successful experiment with a mesoscopic circuit should
therefore be classified as ``non-contextual hidden variable theories''
(NCHV). While a local hidden variable (LHV) theory might still explain the
results, it would have to introduce so far unknown interactions.  Replacing
the belief in locality by the more general {\em assumption} of
non--contextuality is indeed natural once one starts to doubt the validity of
locality in quantum mechanics (see
\cite{Leggett03,Groeblacher07,Paterek07,Branciard07} 
for recent attempts to explain the quantum world with non--local realistic
theories).  The conclusions drawn from a NCHV theory in terms of
bounds on correlations are exactly the same as for a LHV
theory --- only the physical origin of the assumed independence of
Alice's measurement results on Bob's settings is different. The
violation of any of the non--contextuality  
conditions discussed below may be viewed as signature of true quantum
correlations in a mesoscopic circuit.

\subsection{Mesoscopic Circuit}  
\begin{figure}
\includegraphics[width=4cm]{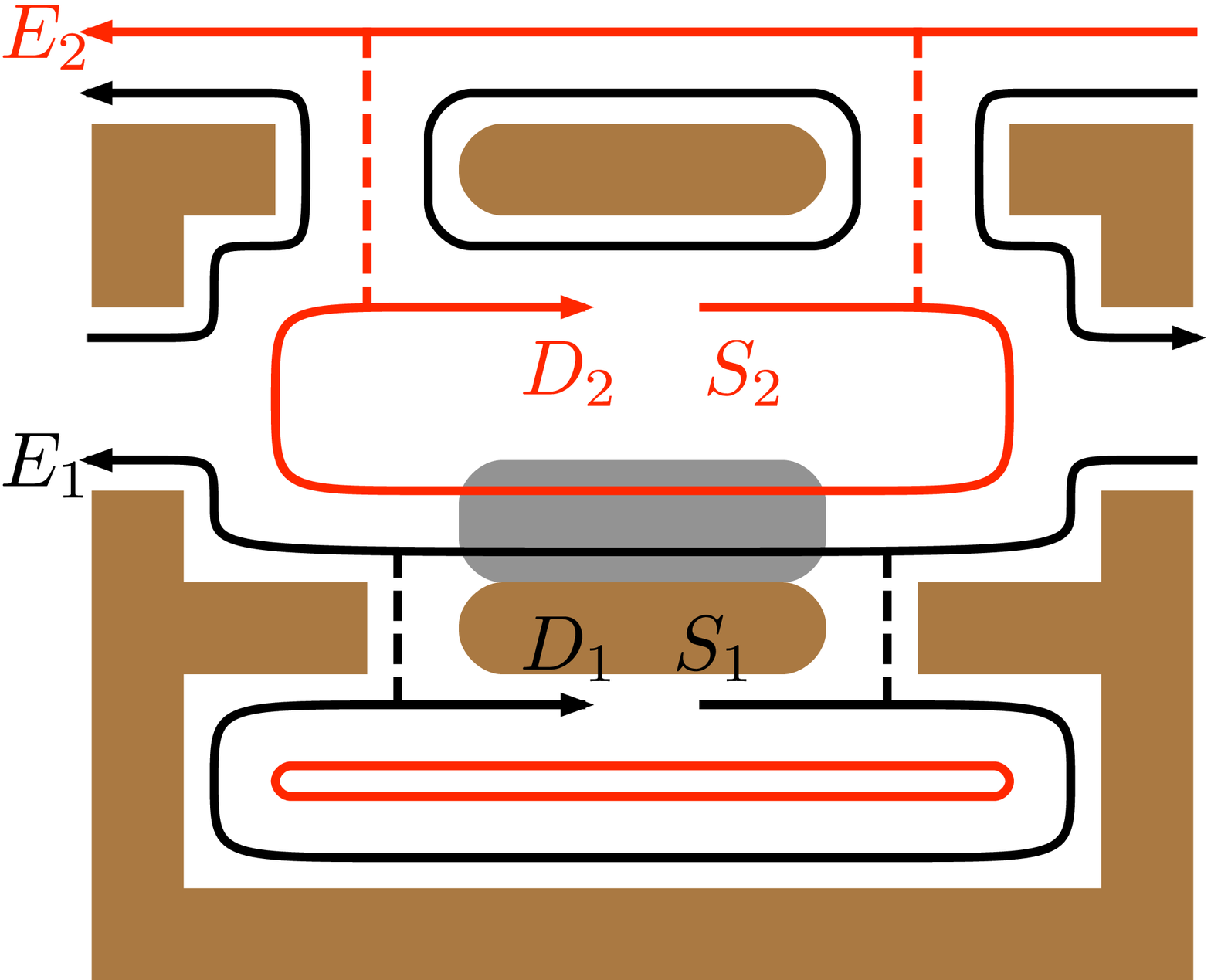}
\includegraphics[width=4cm]{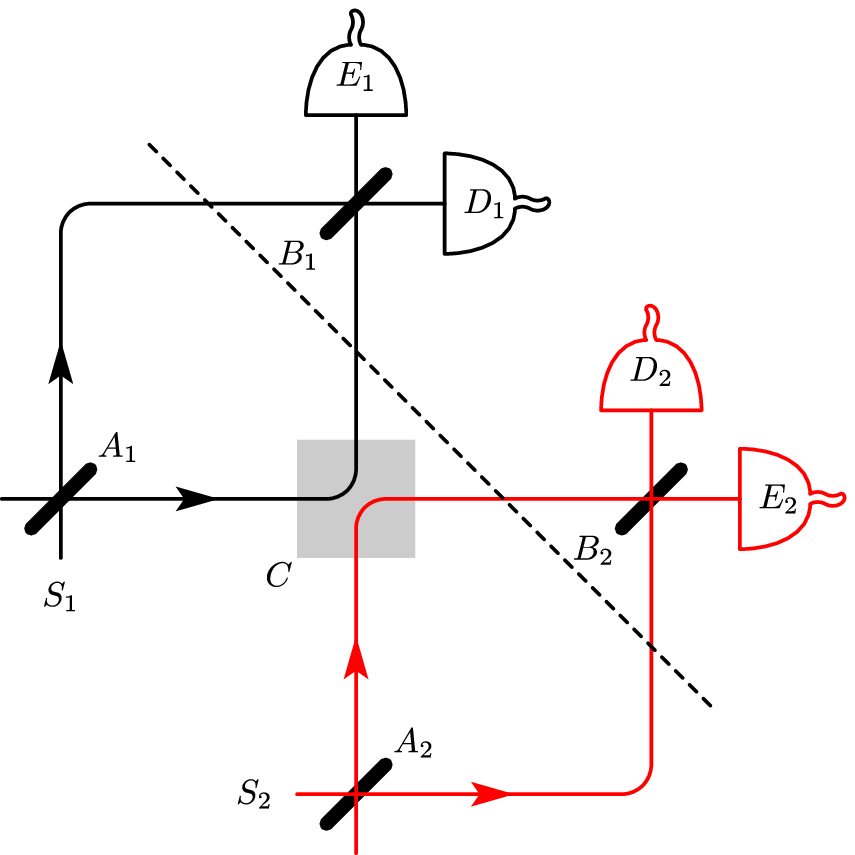}
\caption{(color online) (a) A schematic of the mesoscopic circuit with two
  coupled Mach-Zehnder interferometers fabricated on a quantum Hall bar and (b)
  its equivalent diagram.  The arrowed (black and red) lines are the edge
  channels of the quantum Hall liquid with filling factor 2.  $S_j$ ($j=1,2$)
  denote the electron sources, $A_j$ and $B_j$ the beam splitters (quantum
  point contacts), $D_j$ and $E_j$ the electron detectors, and $C$ the
  conditional phase shift.  The coupling is realized by the Coulomb
  interaction between the two channels in the gray shaded region in (a).}
\label{Hardy::fig:1}
\end{figure}
The circuit consists of two coupled electronic Mach-Zehnder (MZ)
interferometers fabricated on a quantum Hall bar
\cite{Ji03a,Neder06a,Neder07a}, see Fig.~\ref{Hardy::fig:1} (a).  One of the
two MZ interferometers is formed of the outer edge channel of the quantum Hall
liquid with filling factor 2, whereas the other uses the inner channel.  The
electronic beam splitters (BS) in the MZ interferometers are realized by
quantum point contacts (QPC).  A QPC can be fine tuned so as either to block
the inner channel completely and partially transmit the outer channel, or to
entirely transmit the outer channel and partially reflect the inner channel.
In this way a QPC can operate as a BS selectively on one of the two edge
channels.  In order to create an entangled state, the two MZ interferometers
should be coupled.  The coupling in our scheme arises from the Coulomb
repulsion between the electrons in the two parallel edge channels. The
repulsive potential affects the phases of the interacting electrons, and thus
the coupling provides a controlled phase shift between the two MZ
interferometers. This has been recently demonstrated experimentally
\cite{Neder06a}.  As we will explain below, the chip should be fed with
synchronized single-electron sources rather than through more common contact
reservoirs, which inject continuous streams of electrons.
Among several possible methods, we propose the use of the coherent
single--electron source based on a coherent capacitor. Demonstrated
in a recent experiments\cite{Feve07a}, this source allows
well-controlled injection times, 
well-defined energy of the injected electrons, as well as good control of
the input current through the frequency of the pump  (about 180 MHz in
Ref.\cite{Feve07a}).

Let us denote by $S_j$ ($j=1,2$) the input port through which electrons are
injected to the $j$th MZ interferometer, by $A_j$ and
$B_j$ the two beam splitters, and by $D_j$ and $E_j$ the two output ports
where the electrons are 
detected; see Fig.~\ref{Hardy::fig:1} (b).  
We attribute the value +1 to detection in $D_j$ (basis state $|0\rangle$),
and -1 to detection in $E_j$ (basis state $|1\rangle$). 
Using the fact that only relative phases between the two branches in an MZ
interferometer are relevant, we may represent the beam splitters by a real
orthogonal matrix of the form
\begin{equation}
U_B(\theta) =
\begin{pmatrix}
\cos\theta & -\sin\theta \\
\sin\theta & \cos\theta
\end{pmatrix}
\end{equation}
in the basis $\{\ket{0},\ket{1}\}$. Phase shifts within a MZ interferometer
are described by
\begin{equation}
U_P(\phi) =
\begin{pmatrix}
1&0 \\
0 & e^{i\phi}
\end{pmatrix}\,.
\end{equation}
The coupling $C$ reads
\begin{equation}
U_C(\phi) =
\begin{pmatrix}
1 &&&\\ &1&&\\ &&1&\\ &&& e^{i2\phi}
\end{pmatrix}
\end{equation}
in the basis $\{\ket{00},\ket{01},\ket{10},\ket{11}\}$ with $\ket{\sigma_1\sigma_2}\equiv\ket{\sigma_1}\otimes\ket{\sigma_2}$.
$A_1$, $A_2$ and $C$ are used for the
preparation of the entangled state,
while $B_1$ and $B_2$ allow to
select the four different 
quantum measurements.  The two
interacting MZ interferometers realize the unitary transformation 
\begin{equation}
\label{Ut}
U = [U_B(\theta_1')\otimes U_B(\theta_2')]
[U_P(\phi_1)\otimes U_P(\phi_2)]
U_C(\phi)[U_B(\theta_1)\otimes U_B(\theta_2)]\,.
\end{equation}

Even though there  
have been experimental demonstrations of high precision single--electron
detection \cite{Bylander05a,Schoelkopf98a,Gustavsson07a,Fujisawa06a}, none of
them is fast enough for nanosecond time scales.  Therefore, we propose to
use low-frequency current cross--correlations instead
\cite{Samuelsson03a,Samuelsson04a,Chtchelkatchev02a,Beenakker03b}. 
The zero--frequency spectral density $S_{M_1M_2}(H_1,H_2)$ of the
correlations  between two 
drains $H_1,H_2$ (where $H_j\in\{D_j,E_j\}$, $j=1,2$) of the current 
fluctuations $\delta{I}_j(H_j)=I_j(H_j)-\langle{I_j(H_j)}\rangle$, is
defined as  
\begin{equation}
S_{M_1M_2}(H_1,H_2) = \int_{-\infty}^\infty{dt}\;
\left\langle\delta{I}_1(H_1,t)\delta{I}_2(H_2,0)\right\rangle\,.
\end{equation}
To simplify notation, we have suppressed the dependence of $\delta{I}_j$ on the
measurement settings, coded in the subscripts of
$S_{M_1M_2}(H_1,H_2)$ ($M_j=X_j$ or $Y_j$).
If the electron pairs from the synchronized sources $S_1$ and $S_2$ are well
separated in each interferometer (at 180MHz operation the temporal width of
the electron wave-package was below 1 ns for optimal current quantization
in
\cite{Feve07a}),
$S_{M_1M_2}(H_1,H_2)$ is directly related to the joint probabilities of
Eqs.~(\ref{l1}-\ref{l4}) \cite{Samuelsson03a,Samuelsson04a}. For example we
have 
\begin{equation}
\label{Hardy::eq:7}
P({+-}|YX)= \frac{2\pi}{eI_0}S_{YX}(D_1,E_2)\,, 
\end{equation}
and correspondingly for the other joint probabilities in
Eqs.~(\ref{l1})--(\ref{l4}), where $I_0$ is the injection current from the
single--electron sources $S_1$ and $S_2$ ( $\approx{}5\,\mathrm{pA}$ in
\cite{Feve07a}).
The requirement for well separated electrons for the validity of
Eq.~(\ref{Hardy::eq:7}) 
and the need for well--defined interaction phases (and thus simultaneous
arrival of the electrons in the interaction area $C$), motivate the use of
synchronized single electron sources. As a byproduct, the production rate of
electron pairs is precisely known. This is important for a convincing
falsification of any NCHV model, as discussed in Section~\ref{intro} at the
beginning. As Eq.~(\ref{Hardy::eq:7}) makes obvious, the joint probabilities
for our setup are normalized relative to the absolute pump rates $I_0$, and
this allows us to avoid the description by an NCHV model as in
Ref.\cite{Santos92}.

\subsection{Parameters for Hardy's Test}
For Hardy's test, one should create the state (\ref{psiIH}).
In principle two full MZ interferometers and thus four BSs are necessary to do
so. For the choice of measurement operators, one would need two more BSs (one
for Alice, one for Bob). However, since these BSs just perform local unitary
rotations, we can combine the last two BSs on each side, and therefore perform
all experiments for the correlations (\ref{l1})--(\ref{l4}) with just two BSs
per party.  We put the beam splitter $A_1$ into a fixed mode associated with
the unitary matrix $V_1=U_B(\pi/4)$, create $V_2=U_B(\theta_0)$ with $A_2$,
and adjust the coupling to $V_0=U_C(\phi_0)$, where the optimal values
$\theta_0$ and $\phi_0$ are given by
\begin{equation}
\cos(2\theta_0) = \cos(2\phi_0) = 2-\sqrt{5} \,.
\end{equation}
This leads to the entangled state
$\ket\psi=V_0(V_1\otimes V_2)\ket{00}$, 
\begin{equation}
\label{Hardy::eq:1}
\ket{\psi} =
\frac{\cos\theta_0}{\sqrt{2}}\left(\ket{00}+\ket{10}\right)
+ \frac{\sin\theta_0}{\sqrt{2}}\left(\ket{01}+e^{i2\phi_0}\ket{11}\right)\,,
\end{equation}
achieved after the electrons pass through 
$A_1$, $A_2$, and 
undergo the conditional phase shift $C$. 

Following the lines of
Refs.~\cite{Ghirardi06a,Ghirardi06b,Goldstein94a,Hardy93a} we implement the
measurements $X_1$ and $Y_1$ by switching the beam splitter $B_1$ between the
two modes associated with the unitary matrices $U_1=U_B(\pi/4)$ and
$W_1=U_P(2\phi_0)U_B(\pi/4)U_P(-2\phi_0)$, and the beam splitter $B_2$ between
the two modes $U_2=U_B(0)$ and $W_2=U_P(\phi_0)U_B(\chi)U_P(-\phi_0)$, where
$\cot\chi=\tan\theta_0\cos\phi_0$.  In other words, the measurements $X_j$ and
$Y_j$ are given by $X_j=U_j^\dag Z U_j$ and $Y_j=W_j^\dag ZW_j$, respectively,
where $Z=\ket{0}\bra{0}-\ket{1}\bra{1}$. The last phase shifts in $W_1$ and
$W_2$ change the phases of the computational basis states immediately before
detection and do not modify the final probabilities. Omitting them brings the
total unitary transformation to the form (\ref{Ut}) with seven parameters,
whose optimal values we have summarized in Table \ref{tab.params}.  It is then
straightforward to see that the quantum mechanical joint probabilities
$P_\psi$ associated with the state $\ket\psi$ in Eq.~(\ref{Hardy::eq:1})
verify Eqs.~(\ref{l1})--(\ref{l3}), whereas instead of Eq.~(\ref{l4}) we have
Eq.~(\ref{PHid}), i.e.~a violation of the inequality in about 9\% of all
cases.

\subsection{Parameters for CHSH Test}
The singlet state can be created as
\begin{eqnarray} \label{sing}
|\psi_s\rangle&=&\Big[U_B(\frac{\pi}{4})\otimes U_B(0)\Big]
U_{CP}(\frac{\pi}{2})
\Big[U_B(\frac{\pi}{4})\otimes U_B(\frac{\pi}{4})\Big]|00\rangle\,.
\end{eqnarray}
The optimal choice of measurements for a maximal violation of (\ref{CHSHp})
is
\begin{equation}
\begin{split}
X_1&=\sigma_z\,,\\
Y_1&=\sigma_x\,,\\
X_2&=\frac{1}{\sqrt{2}}\left(\sigma_z
  - \sigma_x\right)\,,\\
Y_2&=-\frac{1}{\sqrt{2}}\left(\sigma_x +
  \sigma_z\right) \,,
\end{split}
\end{equation}
where $\sigma_x$ and $\sigma_z$ are Pauli matrices.
This implies that the BS $B_1$ should be set to full transmission ($U_B(0)$)
for $M_1=X_1$, and to $U_B(-\frac{\pi}{4})$ for $M_1=Y_1$, as
$X_1=U_B(\frac{\pi}{4})ZU_B(-\frac{\pi}{4})$. For $B_2$ we choose angles
$\frac{\pi}{8}$ or $\frac{3\pi}{8}$ for $X_2$ or $Y_2$, respectively, as
$X_2=U_B(-\frac{\pi}{8})ZU_B(\frac{\pi}{8})$ and
$Y_2=U_B(-\frac{3\pi}{8})ZU_B(\frac{3\pi}{8})$.
In table \ref{tab.params} we summarize the parameters for the different BSs
and phase shifters for all 4 measurements for Hardy's test and the CHSH test.

\begin{table}[t]
\begin{center}
\begin{tabular}{|c|c|c|c|c|c|c|c|c|}\hline
&$M_1,M_2$ &$\theta_1^{opt}$ &$\theta_2^{opt}$
&$\phi^{opt}$ &$\phi_1^{opt}$ &$\phi_2^{opt}$
&$\theta_1^{'opt}$ &$\theta_2^{'opt}$ \\\hline
Hardy&$X,X$&$\pi/4$&$\theta_0$&$\phi_0$&0&0&$\pi/4$&0\\
Hardy&$X,Y$&$\pi/4$&$\theta_0$&$\phi_0$&0&${-}\phi_0$&$\pi/4$&$\chi$\\
Hardy&$Y,X$&$\pi/4$&$\theta_0$&$\phi_0$&$-2\phi_0$&0&$\pi/4$&0\\
Hardy&$Y,Y$&$\pi/4$&$\theta_0$&$\phi_0$&$-2\phi_0$& ${-}\phi_0$
&$\pi/4$&$\chi$\\  
\hline
CHSH&$X,X$&$\pi/4$&$\pi/4$&$\pi/2$&0&$\pi$&$\pi/4$&$\pi/8$\\
CHSH&$X,Y$&$\pi/4$&$\pi/4$&$\pi/2$&0&$\pi$&$\pi/4$&$3\pi/8$\\
CHSH&$Y,X$&$\pi/4$&$\pi/4$&$\pi/2$&0&$\pi$&$0$&$\pi/8$\\
CHSH&$Y,Y$&$\pi/4$&$\pi/4$&$\pi/2$&0&$\pi$&$0$&$3\pi/8$\\
\hline
\end{tabular}
\caption{Optimal parameters for all four measurements for Hardy's test and
  the CHSH test; $\theta_0=\phi_0=(\arccos(2-\sqrt{5}))/2$,
  $\chi=\mathrm{arccot}(\tan\theta_0\cos\phi_0)$. \label{tab.params}}
\end{center}
\end{table}

\subsection{Imperfections}\label{sec.imp}

Let us now calculate, both for Hardy's test and the CHSH test, the range of
fluctuations of the seven angles $\theta_j$, $\theta_j'$, $\phi_j$, ($j=1,2$),
and $\phi$ of the unitary matrix $U$ in Eq.~(\ref{Ut}) around their optimal
values (see Table \ref{tab.params}), for which QM still predicts a violation
of (\ref{CHSHp}).  Fluctuations in these angles lead to mixed states $\rho$ at
the output ports, which depend on the test and the measurements. We denote the
quantum mechanical predictions of the probabilities corresponding to $\rho$ by
$P_\rho$.  The final QM joint probabilities in (\ref{CHSHp}) are given by
post-processing the $P_\rho$ with a classical stochastic map that models
particle loss.  Besides electrons going undetected, an electron may also be
detected in the wrong output port ($E_j$ instead of $D_j$, or vice versa), or
an electron might be detected erroneously in both output ports.  The last
process, witnessed by a finite value of the correlator $S_{M_1M_2}(E_j,D_j)$,
should be taken care of experimentally by subtracting the dark count.
Detecting an electron in the wrong drain can be shown to be equivalent to a
bit flip error after the state preparation and can be included in the
fluctuations of the 7 angles.  As we exclude joint dark counts, the final
predictions for the joint probabilities in (\ref{CHSHp}) are then simply
obtained by multiplying the $P_\rho$ with a factor $(1-r)^2$, where $r$ is the
single electron loss rate per interferometer, indicating that no electron was
lost, neither on Alice's nor Bob's side. The single particle probabilities are
multiplied only with a factor $(1-r)$. This leads to inequality
(\ref{CHSHr}). 
Note that $r$ can be measured through $\langle
I(D_j)\rangle+\langle I(E_j)\rangle=I_0(1-r)$ for known $I_0$;
see the discussions in Section~\ref{sec:particle-loss}.

\begin{figure}
\centering
\includegraphics[scale=0.25]{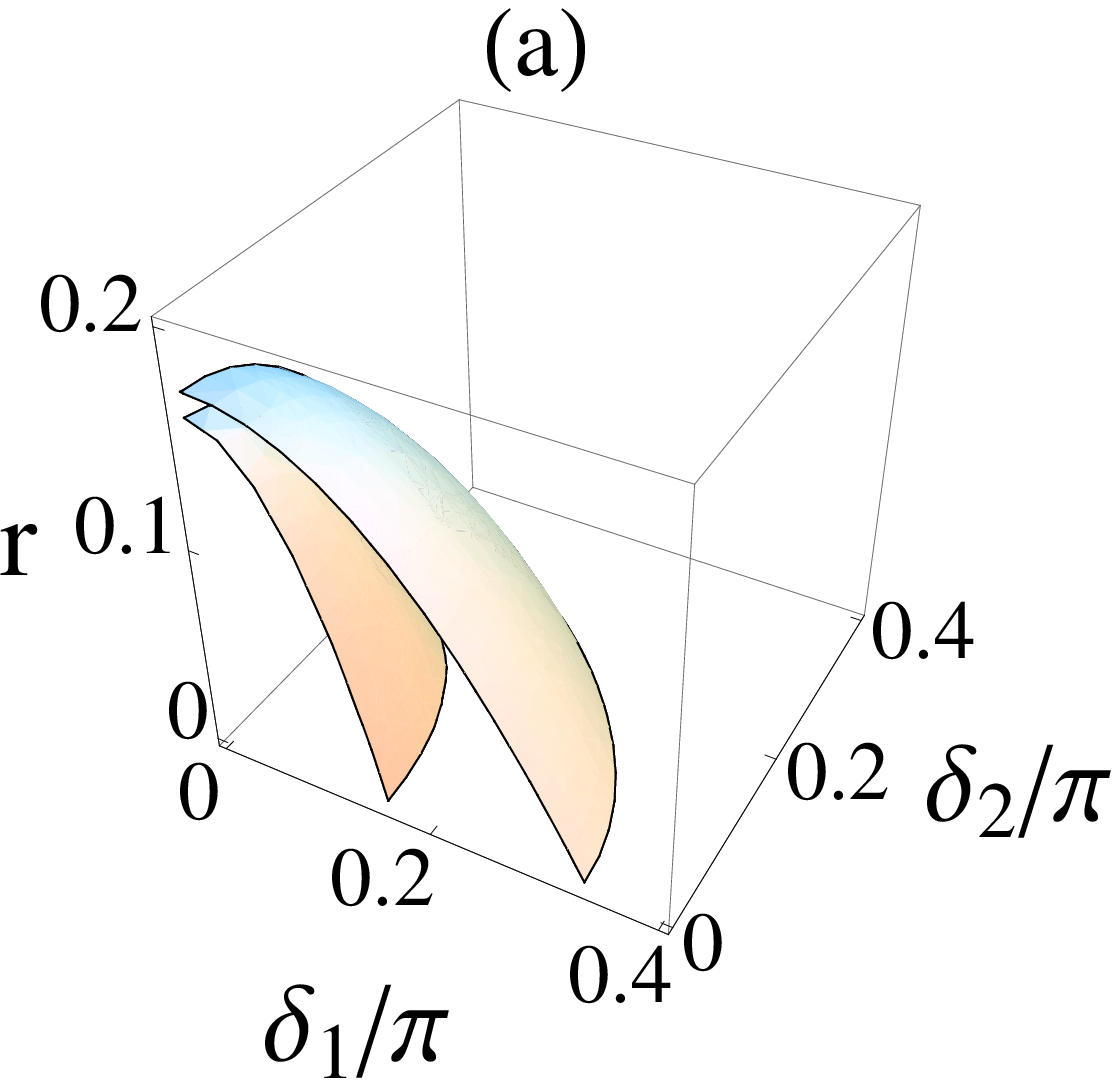}
\includegraphics[scale=0.25]{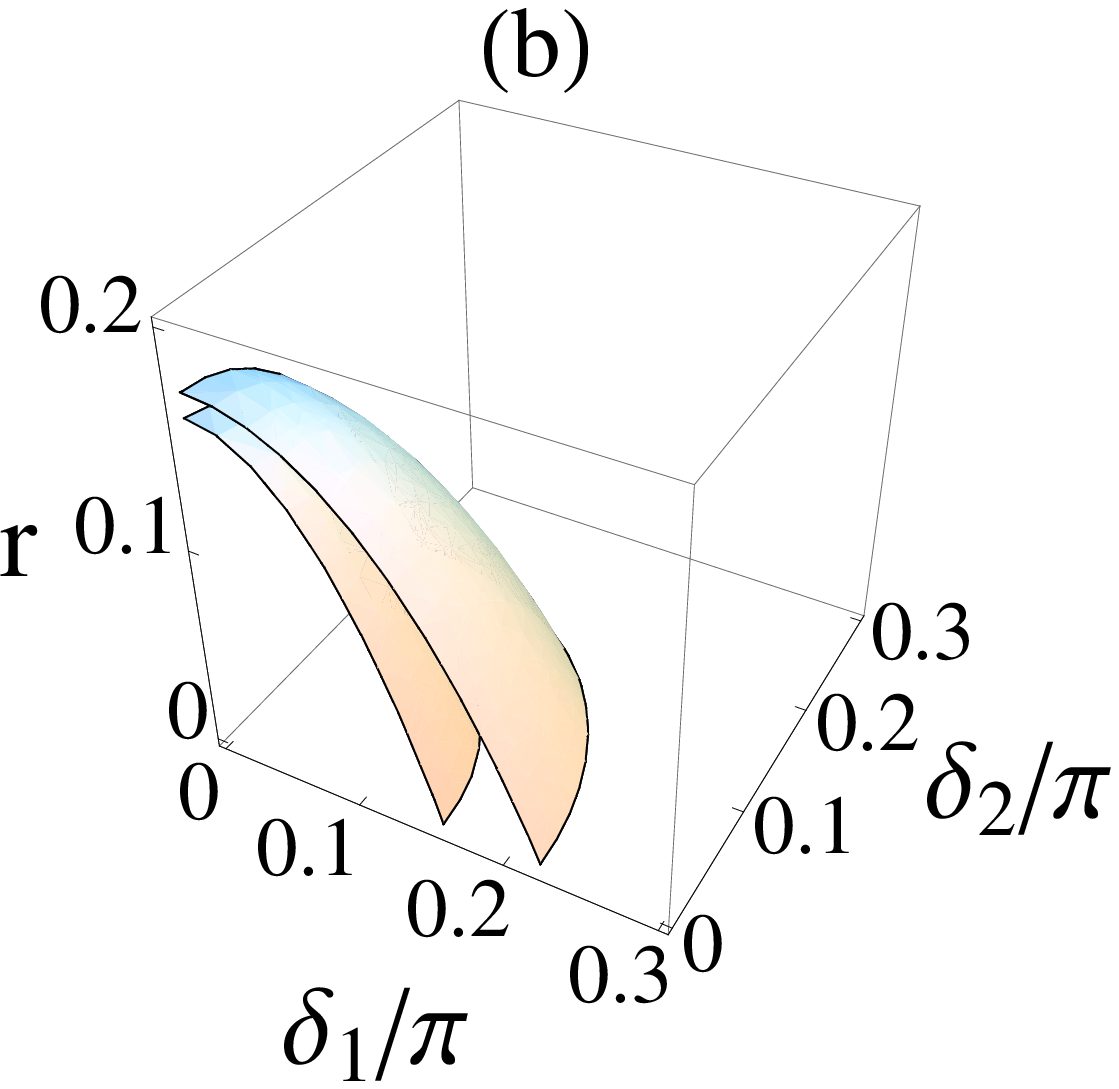} \\
\includegraphics[scale=0.25]{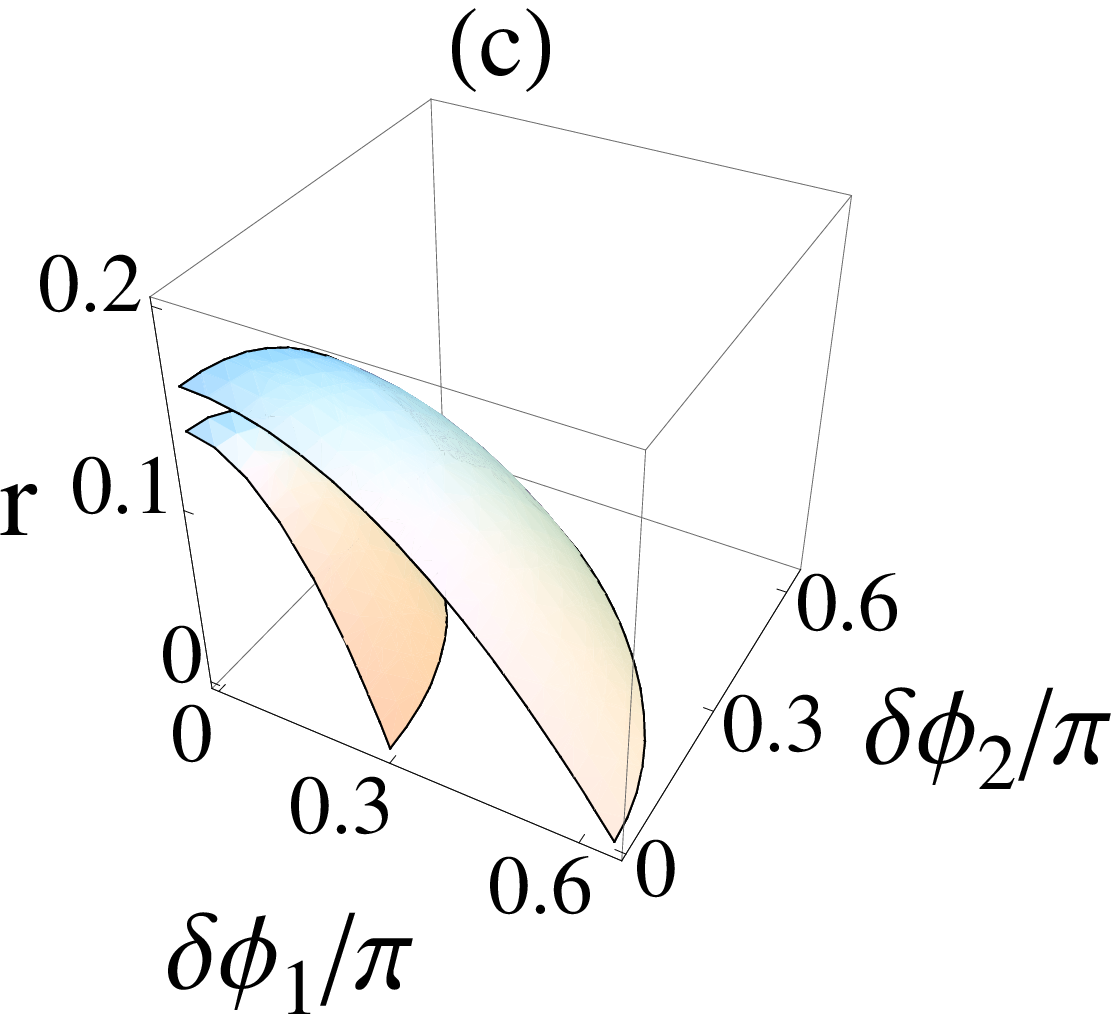}
\includegraphics[scale=0.25]{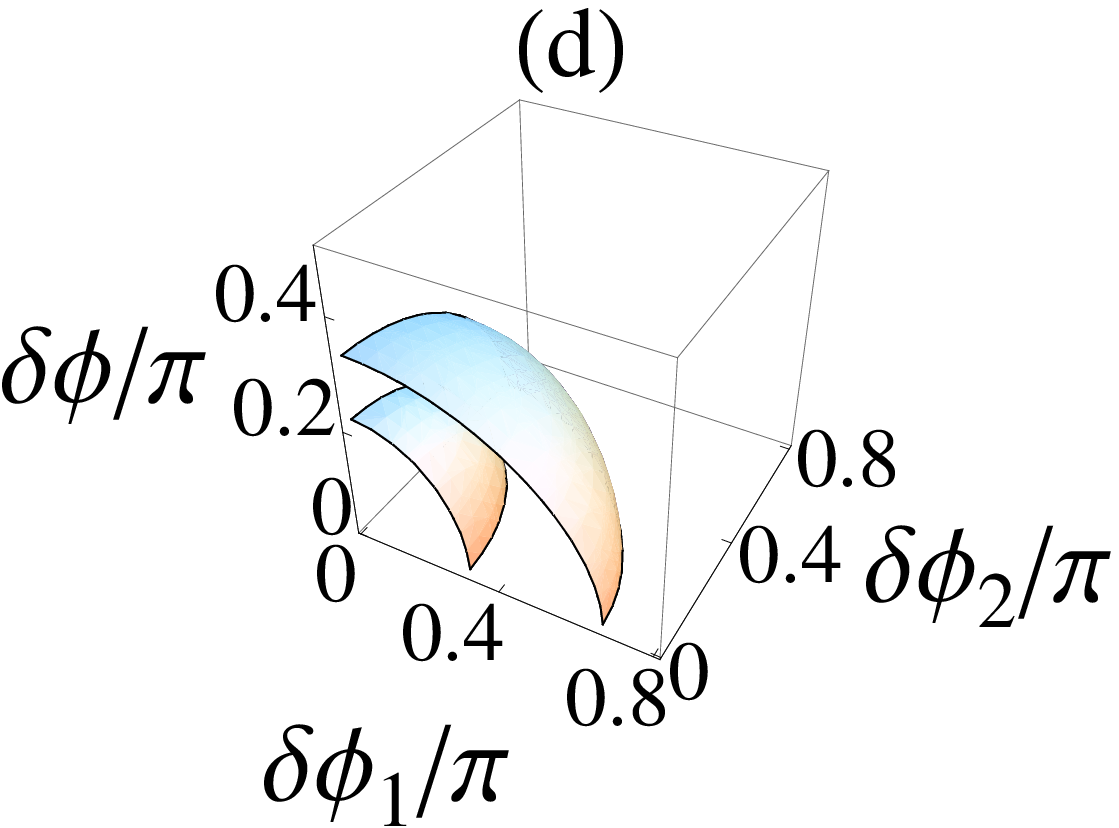}
\caption{Boundaries of the regions where NCHV theories are expected to be
  falsified in Hardy's test (lower surfaces) and the CHSH test (upper
  surfaces) for a selection of different errors.  (a) Allowed fluctuations
  $\delta_1$ and $\delta_2$ in $\theta_1$ and $\theta_2$ of the beam splitters
  $A_1$ and $A_2$, respectively, when electrons are lost with probability $r$
  per interferometer.  (b) The same as (a) for the beam splitters $B_1$ and
  $B_2$.  (c) The same as (a), but for the phase fluctuations $\delta\phi_1$
  and $\delta\phi_2$ in the two Mach--Zehnder interferometers. (d) Same as
  (c), but with a phase fluctuation $\delta\phi$ in the controlled phase shift
  $U_C(\phi)$ instead of $r$ for $r=0$.}
\label{Hardy::fig:2}
\end{figure}

We have estimated the 
allowed range of errors for
uniform distributions of the relevant fluctuations, independent of
the measurement chosen.  Note that different probabilities correspond to
different settings of the BS, and thus to a different final
state before the application of the measurement operator $Z$, even without
fluctuations.  Correspondingly, also a different final 
mixed state is produced for 
each different joint probability function. This can be considered as
Schr\"odinger picture 
of the measurement process (same operator, different states) in
contrast to the more familiar Heisenberg picture (fixed state, different
measurements). Since the CHSH inequality (\ref{CHSHp}) is
linear in the 
probabilities, we may as well average the probabilities themselves over the
distributed parameters.
For example, when the first  beam splitters
$A_1$ and $A_2$
are subject to errors in the tuning of the transmissions, the allowed error
range is given by the condition
\begin{multline}
\label{CHSHpf}
\frac{1}{4\delta_1\delta_2}
\int_{\theta_1^\mathrm{opt}-\delta_1}^{\theta_1^\mathrm{opt}+\delta_1}
{d\theta_1}
\int_{\theta_2^\mathrm{opt}-\delta_2}^{\theta_2^\mathrm{opt}+\delta_2}
{d\theta_2}\;\Bigg\{
\Big[P_\psi({++}|XX)+P_\psi({++}|XY) \\{} +
P_\psi({++}|YX)-P_\psi({++}|YY)\Big](1-r)
-P_\psi^{(1)}({+}|Y)-P_\psi^{(2)}({+}|Y) \Bigg\} \le0
\end{multline}
where $\theta_j^\mathrm{opt}$ ($j=1,2$) are the optimal settings for the
corresponding test and measurements (different for different terms in
the integral, see Table \ref{tab.params}).
In the above inequality~(\ref{CHSHpf}), $P_\psi(m_1,m_2|M_1,M_2)$ denote the
quantum mechanical probabilities corresponding to the states
\begin{math}
|\psi(\theta_1,\ldots,\theta_2')\rangle
=U(\theta_1,\ldots,\theta_2')|00\rangle
\end{math}
with $U$ from Eq.~(\ref{Ut}).  As such, the $\theta_j$-dependence of
$P_\psi(m_1,m_2|M_1,M_2)$ is through the state
$\ket{\psi(\theta_1,\cdots,\theta_2')}$.
Note that the marginal probability $P_\psi^{(1)}({+}|Y)$ is implemented
experimentally either by
\begin{math}
P_\psi^{(1)}= P_\psi({++}|YY)+P_\psi({+-}|YY)
\end{math}
or by
\begin{math}
P_\psi^{(1)}= P_\psi({++}|YX)+P_\psi({+-}|YX)
\end{math},
and  similarly for $P_\psi^{(2)}({+}|Y)$.

Figure~\ref{Hardy::fig:2} shows that for Hardy's test all angles can fluctuate
over intervals of the order of 1 radian, if no electron is lost, whereas for
small fluctuations of the angles loss rates up to 15\% in each interferometer can be tolerated, in 
agreement with earlier findings about the detector efficiency needed to avoid
a detector loophole for Hardy non--locality \cite{Hwang96}.  However, for the
CHSH test the allowed range of fluctuations is even larger (see
Fig.~\ref{Hardy::fig:2}).
Altogether, we see that our mesoscopic scheme is very robust against possible
imperfections. The scheme might therefore be sufficiently robust for
experimental implementation with present day technology.

\section{Conclusions}
We have compared three different tests of quantum non--locality, both
on a theoretical level, and with respect
to a possible implementation in a mesoscopic circuit. We have shown that
Hardy's test becomes a special instance of the more general CHSH inequality
as soon as imperfections have to be taken into account. We have uncovered
the deeper geometrical reason for this fact by   
establishing that Hardy's equations describe a 4D convex polytope embedded 
inside the 
interface, described by a CHSH inequality, between the polytope of 
local correlations and the remaining causal correlations. We
have also 
demonstrated that the inequality $I_{2233}$, relevant for experiments
with three outcomes per observable, is superseeded by the CHSH
inequality if the third measurement outcome is particle loss. We have proposed
a flexible measurement setup based on two interacting Mach--Zehnder
interferometers formed from Hall--bar edge states at filling factor 2 and
quantum point contacts, which allows to implement both Hardy's test and the
CHSH test. Based on that setup, we have shown that both Hardy's test and
the CHSH test should be sufficiently robust with respect to
parameter fluctuations to allow falsification of a non--contextual
hidden variable description of the
experiment. For the CHSH test the tolerance with respect to
fluctuations of the relevant experimental parameters is substantially larger
than for Hardy's test.

\begin{acknowledgments}
D.B.~would like to thank H\'el\`ene Bouchiat, Izhar Neder, and Bertrand Reulet
for interesting discussions. This work was supported in part by the Agence
National de la Recherche (ANR), project INFOSYSQQ, contract number
ANR-05-JCJC-0072, and the EC IST-FET project EUROSQIP.
M.-S.C. was supported by the SRC/ERC program (R11-2000-071), the KRF Grants
KRF-2005-070-C00055 and KRF-2006-312-C00543), the BK21 Project, and the KIAS.
\end{acknowledgments}
% \bibliography{Hardy}
% \bibliography{alison,cience,mathey,physey,conmat,staphy,quaphy,mywork}
\bibliography{Daniel}
\end{document}